\documentclass[11pt,twoside]{article}
\usepackage{./asp2014}

\aspSuppressVolSlug
\resetcounters

\begin{document}

\title{Indirect Detection of Extrasolar Planets via Astrometry}
\author{Brenda C.~Matthews,$^1,^2$ and Bryan Butler,$^3$} 
\affil{$^1$Herzberg Astronomy \& Astrophysics Research Centre, National Research Council of Canada, Victoria, BC, Canada; \email{brenda.matthews@nrc-cnrc.gc.ca}}
\affil{$^2$Department of Physics \& Astronomy, University of Victoria, Victoria, BC, Canada; }
\affil{$^3$National Radio Astronomical Observatory, Socorro, NM, U.S.A. }

\paperauthor{Brenda C.~Matthews}{brenda.matthews@nrc-cnrc.gc.ca}{ORCID_Or_Blank}{National Research Council of Canada}{Herzberg Astronomy \& Astrophysics Research Centre}{Victoria}{BC}{V9E 2E7}{Canada}
\paperauthor{Bryan Butler}{Author2Email@email.edu}{ORCID_Or_Blank}{Author2 Institution}{Author2 Department}{City}{State/Province}{Postal Code}{Country}


\section{Introduction}


There are now thousands of detected and confirmed exoplanets and exoplanetary systems.   Through transit detections, {\it Kepler} has driven home the ubiquity of extrasolar planets and the diversity of extrasolar planetary systems, providing statistics of extrasolar planets in relatively close orbits around their parent stars (Borucki et al. 2010; Howell et al. 2014). The recently launched {\it TESS} (Transiting Exoplanet Survey Satellite) will search for planets around 2 million stars in the solar neighbourhood in the coming two years (Ricker et al. 2015).  The radial velocity technique of detecting the radial velocity variations of the star due to a planet's motion around it is, like the transiting technique, limited to relatively close in planets and biased to planets which are in or near orbits viewed edge-on.  In addition, these methods cannot be utilized toward young or later type stars, which exhibit variable emission from their chromospheres that preclude RV measurements to the required precision.  Direct detection of exoplanets through imaging is a growth industry and a key area of current research.  Optical direct imaging surveys have established that there is not a large population of giant planets on orbits of $>10$ AU in young  ($< 100$ Myr) systems (e.g., Biller \& Close 2007; Nielsen 2011; Galicher et al. 2016). Older systems are not accessible to direct imaging because the thermal emission of the planets must be detected and this is only strong enough for the current generation of instrumentation for young planets.  

\begin{figure}
\begin{center}
 \includegraphics[width=0.5\textwidth]{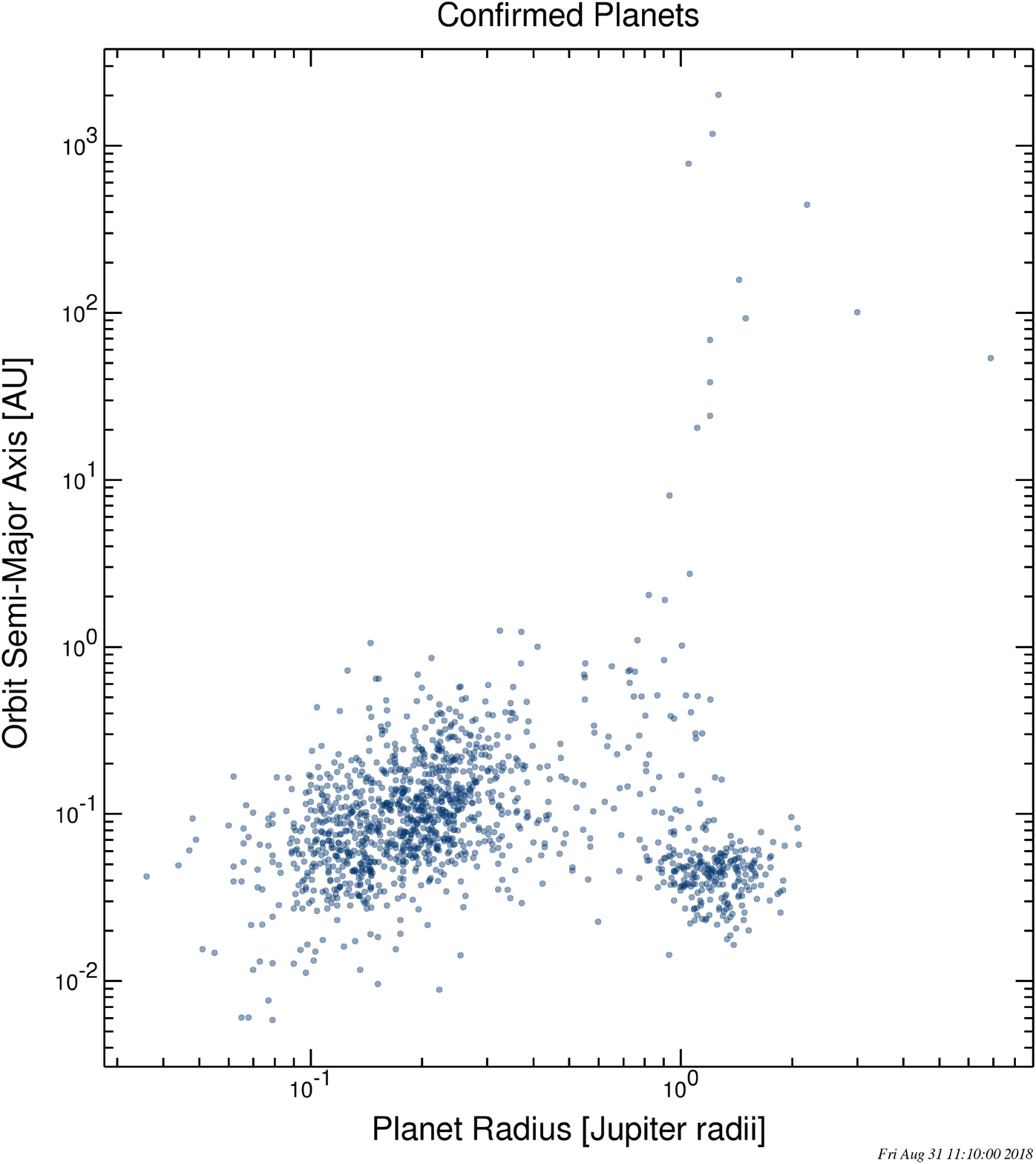}
\caption{Known exopolanets, as of August 31, 2018 (NASA Exoplanet Archive web page).}
\label{known}
\end{center}
\end{figure}

The population of planets in systems of the age and on the scales of the outer solar system therefore remains uncharacterized.   This region of the parameter space for host stars can be targeted for planet searches using the astrometric capabilities of the ngVLA.  The idea of detection of extrasolar planets through astrometry in the radio is not a new one (Butler, Wootten, \& Brown 2003), but ngVLA will have enhanced capabilities compared to the current VLA or ALMA due to its longer baselines.  The ngVLA will have the sensitivity to detect extrasolar planets via their induced motions onto their parent stars, an ``indirect detection'' method similar to the RV technique. By monitoring the parent star, it is possible to infer the reflex motion of the planet. Because this technique measures plane of sky motion, it is complimentary to the RV and transit (Kepler) methods of such detections, and so probes an orthogonal parameter space.  Figure~\ref{known} shows the currently confirmed exoplanets as of August 31, 2018; 3778 in total (from the NASA Exoplanet Archive web page).  Note the paucity of planets in orbits similar to our own giant planets in this plot (those with orbits $>$ 5 AU) - this is a well-known conundrum in exoplanet studies.  Astrometric techniques are particularly well-suited to search for just such planets, since more massive planets in orbits further from their primary induce a larger reflex motion of that primary.

With so many planets known, what is the added impact of characterization of outer solar systems? Much of the dynamical evolution of our own solar system was driven by the outer planets. Indeed the Nice model is dependent on a violent migration event (Morbidelli et al. 2007). Even more quiescent theories of evolution depend on the ice giant migration, particularly Neptune to corral the outer planetesimal belt and stabilize the orbits of the giant planets. 

There is now evidence of significant substructure at large orbital radii in protoplanetary disks (e.g., HL Tau, Elias 2-27 etc.) which suggests that formation mechanisms of planets may be active quite distant from the star, at least at early times. While our own solar system's size is typically given as the radius of the Kuiper Belt (i.e., not much beyond Neptune; Sibthorpe et al. 2018), many other debris disks (dusty circumstellar disks created by extrasolar planetesimal belts) show much larger radial extent, suggesting exoplanets may also exist at larger radii in such systems. We can search for evidence of wide orbit Jupiter and Neptune analogues using the ngVLA. 




Ground based efforts to find extrasolar planets have focused in recent years in several areas, including differential astrometry, radial velocity measurements, and gravitational lensing. There are also proposals to directly image Jovian-class planets near a small number of stars. Planet detection using astrometry with ngVLA will both complement these efforts and have several important advantages:

%
%
%

\begin{itemize}
\item{Astrometric searches for planets with ngVLA will use absolute (wide-angle) astrometry, avoiding the problem of solving for motions in both a target star and a background reference star. ngVLA observations will be able to tie stellar positions directly to the quasar reference frame, since both stars and quasars will be bright enough for high-accuracy astrometry with ngVLA;}
\item{Systematic errors will be lower with ngVLA than with techniques at optical or IR wavelengths because ngVLA can observe stars 24 hours a day, providing various types of closure constraints on ngVLA astrometry. This will reduce the kinds of seasonal systematic errors that plague optical astrometry. The ngVLA at 10-100 GHz will also have less issues with weather constraints than ALMA will at higher frequencies;}
\item{Astrometry measures the position of the star, so that the mass of the unseen planet can be easily determined. Direct detection techniques, either ground or space-based, cannot determine planetary masses, but will provide constraints on the parameters of a particular planetary system that complement those provided by astrometry; and}
\item{Astrometric results from ngVLA will directly complement the results expected from differential astrometry with proposed ground-based IR interferometers and ALMA. ngVLA will not face the constraints on finding suitable background reference stars that are required by differential IR astrometry, so some additional stellar systems may be accessible to ngVLA compared to the optical/IR.}
\end{itemize}

\section{Astrometry of planet hosts}

The orbit of any planet around its central star causes that star to undergo a reflexive motion around the star-planet barycenter. Figure~\ref{solarTB} shows the orbit of the Sun on the planet of the sky if it were viewed from 30 light years ($\sim$9 pc).  Those motions are of order milliarcseconds over periods of tens of years - it is these motions that can be astrometrically detected by ngVLA.  By taking advantage of the high resolution provided by the ngVLA at even intermediate baselines, we will be able to detect this motion and thereby constrain the planetary population at large orbital radii for many solar type stars.  We note that this technique has been used for decades to determine the orbits of multiple star systems (e.g., Ransom et al. 2012).

\begin{figure}
\begin{center}
 \includegraphics[width=0.6\textwidth]{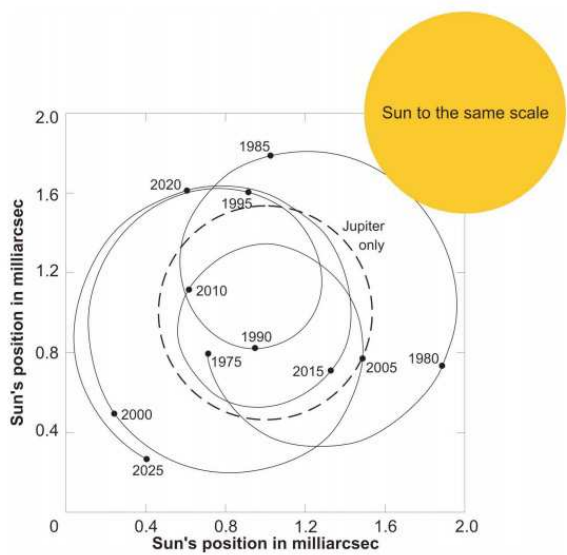}
\caption{The wobble of the Sun on the plane of the sky, if it were viewed face on from 30 light years distance (Jones 2008).}
\label{solarTB}
\end{center}
\end{figure}

Making the usual approximation that the planet mass is small compared to the stellar mass, the stellar orbit projected on the sky is an ellipse with angular semi-major axis $\theta_r$ (in arcsec) given by the product of the ratio of the planetary mass ($M_p$) to stellar mass ($M_*$) and the ratio of the semi-major axis of the planet in AU ($a_{AU}$) and the distance to the star in pc ($d_{pc}$):
\begin{equation}
\theta_r = \frac{M_p}{M_*} \ \frac{a_{AU}}{d_{pc}} \quad .
\end{equation}

The astrometric resolution of an interferometer, the angular scale over which changes can be discriminated ($\Theta$), is proportional to the intrinsic resolution, $\theta_{HPBW}$, and inversely proportional to the signal to noise with which the stellar flux density is detected ($SNR^*$):
\begin{equation}
\Theta = \frac{\theta_{HPBW}}{2 \times SNR^*} \quad .
\end{equation}
Note that there are systematic effects which limit this at high SNR (Reid \& Honma 2014), but we are not expecting to be in that regime for these kinds of observations.

In order to detect the wobble of a star caused by an orbiting planet we must therefore calculate the induced motion on the star (meaning we must know distance to the system, the mass of the star and the assumed mass and orbit of the planet), we must know the resolution of the ngVLA, and we must know the SNR (the flux density of the star, and the sensitivity of ngVLA).  We assume we can determine the distance to the system and mass of the star from information in a catalog.  We also assume that we can determine effective temperature either directly or indirectly from that information.  Note that at wavelengths shorter than about 2 cm, the emission from stars is mostly thermal emission from the disk of the star (see Figure~\ref{solarTB}, which shows this for the Sun).  If we assume that the brightness temperature of the star is just the effective temperature, then we can calculate the flux density of the star via:
\begin{equation}
   F_* = {{2 \, h \, \nu^3} \over {c^2}} \,
           {1 \over {e^{h \, \nu / k \, T_{eff}} - 1}} \Omega \quad ,
\end{equation}
where $h$ is the Planck constant, $c$ is the speed of light in vacuum, $k$ is the Boltzmann constant, $T_{eff}$ is the effective temperature of the planet, and $\Omega$ is the angular size of the star.  We assume the diameter of the star can be determined from the information in the catalog.  To calculate the sensitivity of ngVLA, we assume we will observe at 30 GHz, and take the noise as 1.5 $\mu$Jy/min (Butler et al. 2018) and assume an on-source integration time of one hour, so final noise of $\sim$0.2 $\mu$Jy.  We assume we need an SNR of 4 for each observation.  We assume a maximum baseline for ngVLA of 800 km, so $\theta_{HPBW} \sim 3$ masec.  For candidate planets, we assume an orbit of 5 AU (since we are interested in solar system giant planet analogs), and assume three masses: 5 times Jupiter, Jupiter, and Neptune mass.

\begin{figure}
\begin{center}
 \includegraphics[width=0.8\textwidth]{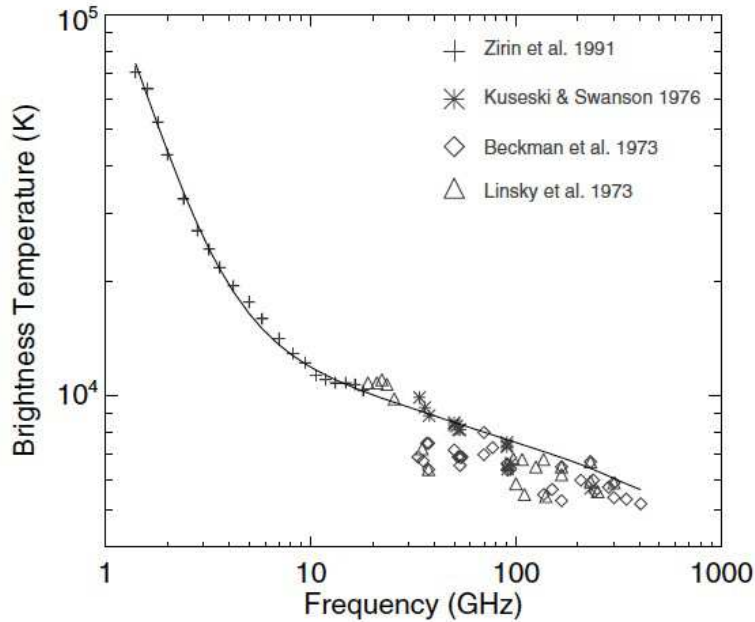}
\caption{Figure showing the solar brightness temperature as a function of wavelength (Selhorst, Silva, \& Costa 2005).}
\label{solarTB}
\end{center}
\end{figure}

We have access to two catalogs of stars that we can use to investigate how many of them may have planets which could be detected in this way: Hipparcos (Perryman et al. 2017) and Gaia (Gaia Collaboration et al. 2016).  Both catalogs have well-determined distances to stars.  The Gaia catalog has radii and effective temperatures for a number of stars.  The Hipparcos catalog does not, but does contain information on spectral type and luminosity class.  We use that information to infer the mass, radii, and effective temperatures using Table 15.8 in Allen's Astrophysical Quantities.  For the Gaia catalog, we determine the mass of the star using the luminosity in the catalog, and the usual mass-luminosity relation.

For the Hipparcos catalog, we find 831 5*Jupiters that can be detected, 538 Jupiters, and 31 Neptunes.  Since that catalog also contains values for B-V, we can classify 'solar analog' stars as those with B-V between 0.62 and 0.71, and with that classification, we can find 5*Jupiters around 43 solar analog stars, Jupiters around 26 solar analog stars, and Neptunes around 2 solar analog stars.

For the Gaia catalog, we find 368 5*Jupiters that can be detected, 68 Jupiters, and 3 Neptunes.  That catalog does not have spectral type or luminosity class information in it, so we define a different type of 'solar candidate' which are stars that have within 30\% of the radius and effective temperature of the Sun.  With that classification we can find 5*Jupiters around 85 solar candidate stars, Jupiters around 10 solar candidate stars, and Neptunes around no solar candidate stars.

There is significant overlap between the two catalogs, of course, and we attribute the difference in the detectable number of stars to be due to the different methods of calculation for the two.  The takeaway is that of order 1000 5*Jupiters can be detected, and 100 Jupiters can be detected, and of these some tens of the stars are similar to the Sun.



\section{Finding planets in practice}

Naturally, the detection of a planet's influence on the star requires regular observations of the star in addition to the interferometric resolution to reasonably detect a gas or ice giant akin to those in the Solar System. We will require detections of the star with a cadence that will allow us to measure the star's position 3 or 4 times during the planet's orbit. We will also have to calibrate for the stellar proper motion. Gaia has now measured the proper motions to high accuracy for many stars in the solar neighbourhood. Selecting out those within 100 pc, i.e., with parallaxes > 10 mas, that catalog has thousands that have proper motions that are known to better than 0.2 mas/yr.  

One can readily see an observational strategy developing in which selected subsamples of stars are routinely observed over a period of several years, and those with promising planetary like variations after 2 observations are then the focus of subsequent observational campaigns.

\section{Uniqueness of ngVLA Capabilities}

ALMA has been highlighted for its capability in detecting extrasolar planets through astrometry in Band 7 (Butler, Wootten, \& Brown 2003). This is akin to the science that can be done at 1 cm (30 GHz) with the ngVLA, except that the ngVLA will have a much better intrinsic resolution, meaning more systems should be detectable under a planetary wobble with ngVLA than with ALMA.  No other facility has such a capability.

\section{Complementarity with other planet search facilities beyond 2025}
	

Interferometers have the resolution needed to do astrometry of nearby stellar systems to the precision needed to detect extrasolar planets.  OST will not have the resolution to do astrometry or detect planets through direct imaging. A suite of upcoming telescopes has various capabilities to directly detect extrasolar planets in scattered light, namely TMT and WFIRST, although the timelines for both are uncertain.  Whether LUVOIR or HABEX are viable in 2025 is hard to foresee. 
 
\ 

\acknowledgements We appreciate comments from Ed Fomalont. 



\ 

\

\noindent {\bf References}

\ 

\noindent Biller, B.A., \& L.M. Close 2007, ApJ, 669, L41. \hfil\break
Borucki, W.J., et al. 2010, Science, 327, 977. \hfil\break
Butler, B., A. Wootten, \& B. Brown 2003, ALMA Memo 475. \hfil\break
Butler, B., W. Grammer, R. Selina, et al. 2018, BAAS, 231, 342.09. \hfil\break
Gaia Collaboration, T. Prusti, J.H.J. de Bruijne, et al. 2016, A\&A, 595, A1. \hfil\break
Galicher, R., et al. 2016, A\&A, 594, id. A63. \hfil\break
Howell, S.B., et al. 2014, PASP, 938, 398. \hfil\break
Jones, B.W. 2008, Int. J. Astrobio., 7, 279. \hfil\break
Morbidelli, A., K. Tsiganis, A. Crida, et al. 2007, ApJ, 134, 1790. \hfil\break
Nielsen, E.L. 2011, PhD Thesis. \hfil\break
Perryman, M.A.C., L. Lindegren, J. Kovalevsky, et al. 1997, A\&A, 323, L49. \hfil\break
Ransom, R.R., N. Bartel, M.F. Bietenholz, et al. 2012, ApJSS, 201, 6.
Reid, M.J., \& M. Honma 2014, ARAA, 52, 339. \hfil\break
Ricker, G.R., et al. 2015, JATIS, 1, id. 014003. \hfil\break
Selhorst, C.L., A.V.R. Silva, and J.E.R. Costa 2005, A\&A, 433, 365. \hfil\break
Sibthorpe, B., G.M. Kennedy, M.C. Wyatt, et al. 2018, MNRAS, 475, 3046. \hfil\break


\end{document}